\documentclass[letterpaper]{article}
\usepackage{natbib,alifeconf}
\usepackage{amsmath}
\usepackage{graphicx,wrapfig,hyperref}

\newcommand{\fig}{Fig.~}

\title{Environmental variability and network structure determine the optimal plasticity mechanisms in embodied agents
}
\author{Emmanouil Giannakakis$^{1,2}$, Sina Khajehabdollahi
$^{1}$ \and Anna Levina$^{1, 2, 3}$ \\
\mbox{}\\
$^1$Department of Computer Science, University of T\"ubingen, T\"ubingen, Germany\\
$^2$Max Planck Institute for Biological Cybernetics, T\"ubingen, Germany\\
$^3$Bernstein Center for Computational Neuroscience  T\"ubingen, T\"ubingen, Germany\\
\bigskip
 }

\begin{document}

\maketitle
\begin{abstract}
The evolutionary balance between innate and learned behaviors is highly intricate, and different organisms have found different solutions to this problem. We hypothesize that the emergence and exact form of learning behaviors is naturally connected with the statistics of environmental fluctuations and tasks an organism needs to solve. Here, we study how different aspects of simulated environments shape an evolved synaptic plasticity rule in static and moving artificial agents. We demonstrate that environmental fluctuation and uncertainty control the reliance of artificial organisms on plasticity. 
Interestingly, the form of the emerging plasticity rule is additionally determined by the details of the task the artificial organisms are aiming to solve. Moreover, we show that co-evolution between static connectivity and interacting plasticity mechanisms in distinct sub-networks changes the function and form of the emerging plasticity rules in embodied agents performing a foraging task.\end{abstract}

\section{Introduction}

One of the defining features of living organisms is their ability to adapt to their environment and incorporate new information to modify their behavior. It is unclear how the ability to learn first evolved \cite{Papini2012}, but its utility appears evident. Natural environments are too complex for all the necessary information to be hardcoded genetically \cite{SnellRood2013} and more importantly, they keep changing during an organism’s lifetime in ways that cannot be anticipated \cite{Ellefsen2014, Dunlap2016}. The link between learning and environmental uncertainty and fluctuation has been extensively demonstrated in both natural \cite{Kerr2003, snellrood2019}, and artificial environments \cite{Nolfi1996}. 

Nevertheless, the ability to learn does not come without costs. For the capacity to learn to be beneficial in evolutionary terms, a costly nurturing period is often required, a phenomenon observed in both biological \cite{Thornton2011}, and artificial organisms \cite{Eskridge2012}. Additionally, it has been shown that in some complex environments, hardcoded behaviors may be superior to learned ones given limits in the agent's lifetime and environmental uncertainty \cite{Dunlap2009, Fawcett2012, Tjarko2020}.

The theoretical investigation of the optimal balance between learned and innate behaviors in natural and artificial systems goes back several decades. However, it has recently found also a wide range of applications in applied AI systems \cite{Lee2020, Biesialska_2020}.
Most AI systems are trained for specific tasks, and have no need for modification after their training has been completed. Still, technological advances and the necessity to solve broad families of tasks make discussions about life-like AI systems relevant to a wide range of potential application areas. 
Thus the idea of open-ended AI agents \cite{deepmind2021} that can continually interact with and adapt to changing environments has become particularly appealing.

Many different approaches for introducing lifelong learning in artificial agents have been proposed. 
Some of them draw direct inspiration from actual biological systems \cite{schmidhuber1987, Parisi2019}. Among them, the most biologically plausible solution is to equip artificial neural networks with some local neural plasticity \cite{ Thangarasa2020}, similar to the large variety of synaptic plasticity mechanisms \cite{Citri2008, Feldman2009, Caroni2012} that performs the bulk of the learning in the brains of living organisms \cite{Magee2020}. The artificial plasticity mechanisms can be optimized to modify the connectivity of the artificial neural networks toward solving a particular task. The optimization can use a variety of approaches, most commonly evolutionary computation.  

The idea of meta-learning or optimizing synaptic plasticity rules to perform specific functions has been recently established as an engineering tool that can compete with state-of-the-art machine learning algorithms on various complex tasks \cite{Burms2015, Najarro2020, Pedersen2021, Yaman2021}.
Additionally, it can be used to reverse engineer actual plasticity mechanisms found in biological neural networks and uncover their functions \cite{Confavreux2020, Jordan2021}.

Here, we study the effect that different factors (environmental fluctuation and reliability, task complexity) 
have on the form of evolved functional reward-modulated plasticity rules. We investigate the evolution of plasticity rules in static, single-layer simple networks. Then we increase the complexity by switching to moving agents performing a complex foraging task. In both cases, we study the impact of different environmental parameters  on the form of the evolved plasticity mechanisms and the interaction of learned and static network connectivity. Interestingly, we find that different environmental conditions and different combinations of static and plastic connectivity have a very large impact on the resulting plasticity rules.

\section{Methods}
\subsection{Environment}
We imagine an agent who must forage to survive in an environment presenting various types of complex food particles. Each food particle is composed of various amounts and combinations of $N$ ingredients that can have positive (food) or negative (poison) values.  The value of a food particle is a weighted sum of its ingredients. To predict the reward value of a given resource, the agent must learn the values of these ingredients by interacting with the environment. The priors could be generated by genetic memory, but the exact values are subject to change.

To introduce environmental variability, we stochastically change the values of the ingredients. 
More precisely, we define two ingredient-value distributions $E_1$ and $E_2$ \cite{Guttenberg2019} and switch between them, with  probability $p_{tr}$ for every time step. 
We control how (dis)similar the environments are by parametrically setting $E_2 = (1 - 2d_e)E_1$, with $d_e \in [0,1]$ serving as a distance proxy for the environments; when $d_e = 0$, the environment remains unchanged, and when $d_e = 1$ the value of each ingredient fully reverses when the environmental transition happens. For simplicity, we take values of the ingredients in $E_1$ equally spaced between -1 and 1 (for the visualization, see \fig\ref{fig:result_static}a, b).

\subsection{Static agent}
\begin{figure}[!htb]
\begin{center}
\includegraphics[width=0.9\columnwidth]{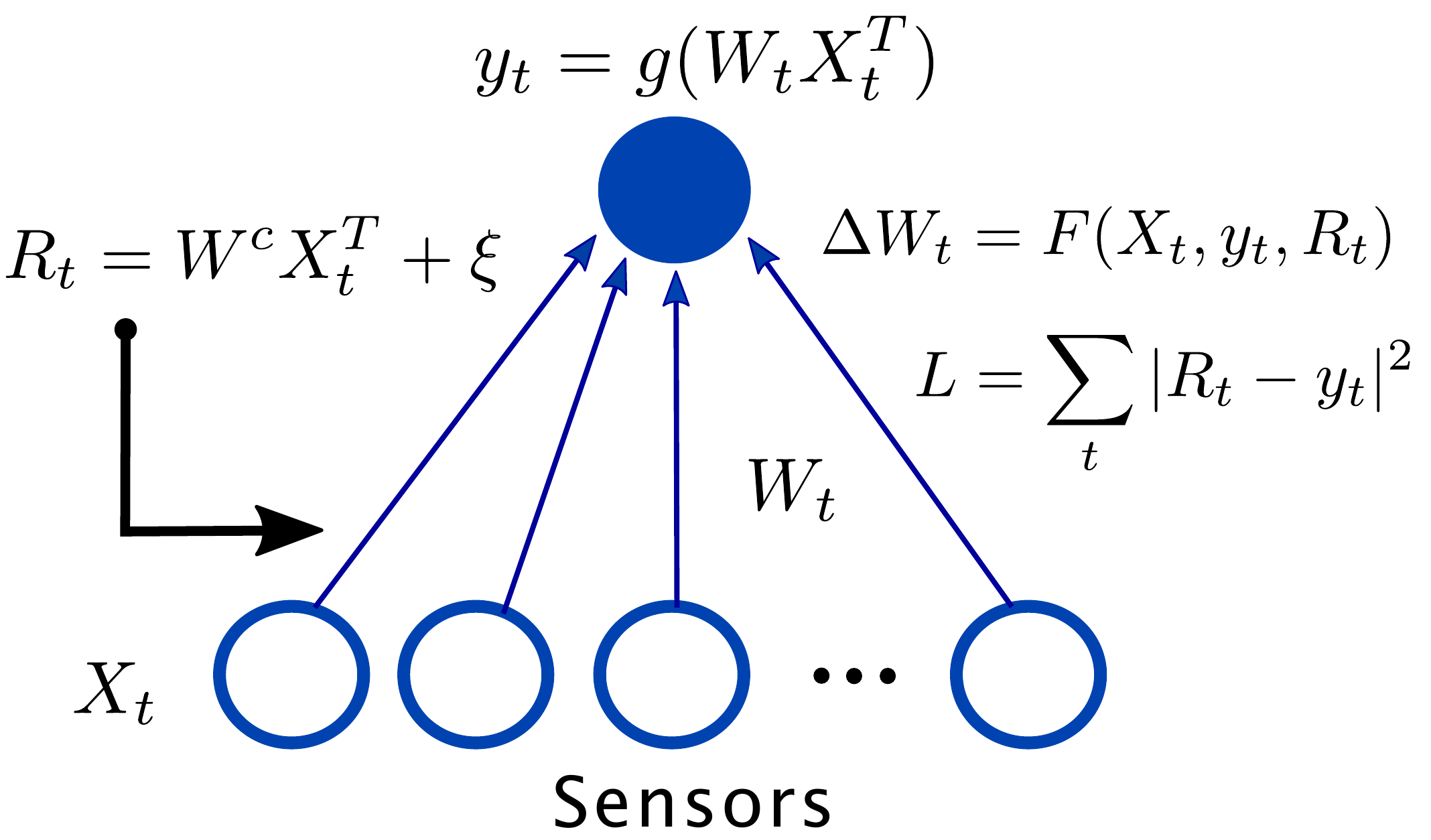}
\caption{\textit{An outline of the static agent's network. The sensor layer receives inputs representing the quantity of each ingredient of a given food at each time step. The agent computes the prediction of the food's value $y_t$ and is then given the true value $R_t$; it finally uses this information in the plasticity rule to update the weight matrix.}}
\label{fig:sensor_network}
\end{center}
\end{figure}

The static agent receives  passively presented food as a vector of ingredients and can assess its compound value using the linear summation of its sensors with the (learned or evolved) weights, see \fig\ref{fig:sensor_network}.
The network consists of $N$ sensory neurons that are projecting to a single post-synaptic neuron.  At each time step, an input $X_t = (x_1, \dots, x_N)$ is presented, were the value $x_i, \ i \in \{1, \dots, N\}$ represents the quantity of the  ingredient $i$. We draw $x_i$ independently form a uniform distribution on the $[0,1]$ interval ($x_i \sim U(0, 1)$). The value of each ingredient $w_i^c$ is determined by the environment ($E_1$ or $E_2$). 

The postsynaptic neuron outputs a prediction of the food $X_t$ value as $y_t = g(W X_t^T)$. Throughout the paper, $g$ will be either the identity function, in which case the prediction neuron is linear, or a step-function; however, it could be any other nonlinearity, such as a sigmoid or ReLU. After outputting the prediction, the neuron receives feedback in the form of the real value of the input $R_t$. The real value is computed as $R_t = W^c X_t^T + \xi,$ where  $W^c = (w_1^c, \dots, w_N^c)$ is the actual value of the ingredients, and $\xi$ is a term summarizing the noise of reward and sensing system $\xi  \sim  \mathcal{N}(0, \sigma)$. 

For the evolutionary adjustment of the agent's parameters, the loss of the static agent is the sum of the mean squared errors (MSE) between its prediction $y_t$ and the reward $R_t$ over the lifetime of the agent. The agent's initial weights are set to the average of the two ingredient value distributions, which is the optimal initial value for the case of symmetric switching of environments that we consider here.

\subsection{Moving Agent}
\begin{figure}[!htb]
\begin{center}
\includegraphics[width= 0.85\columnwidth]{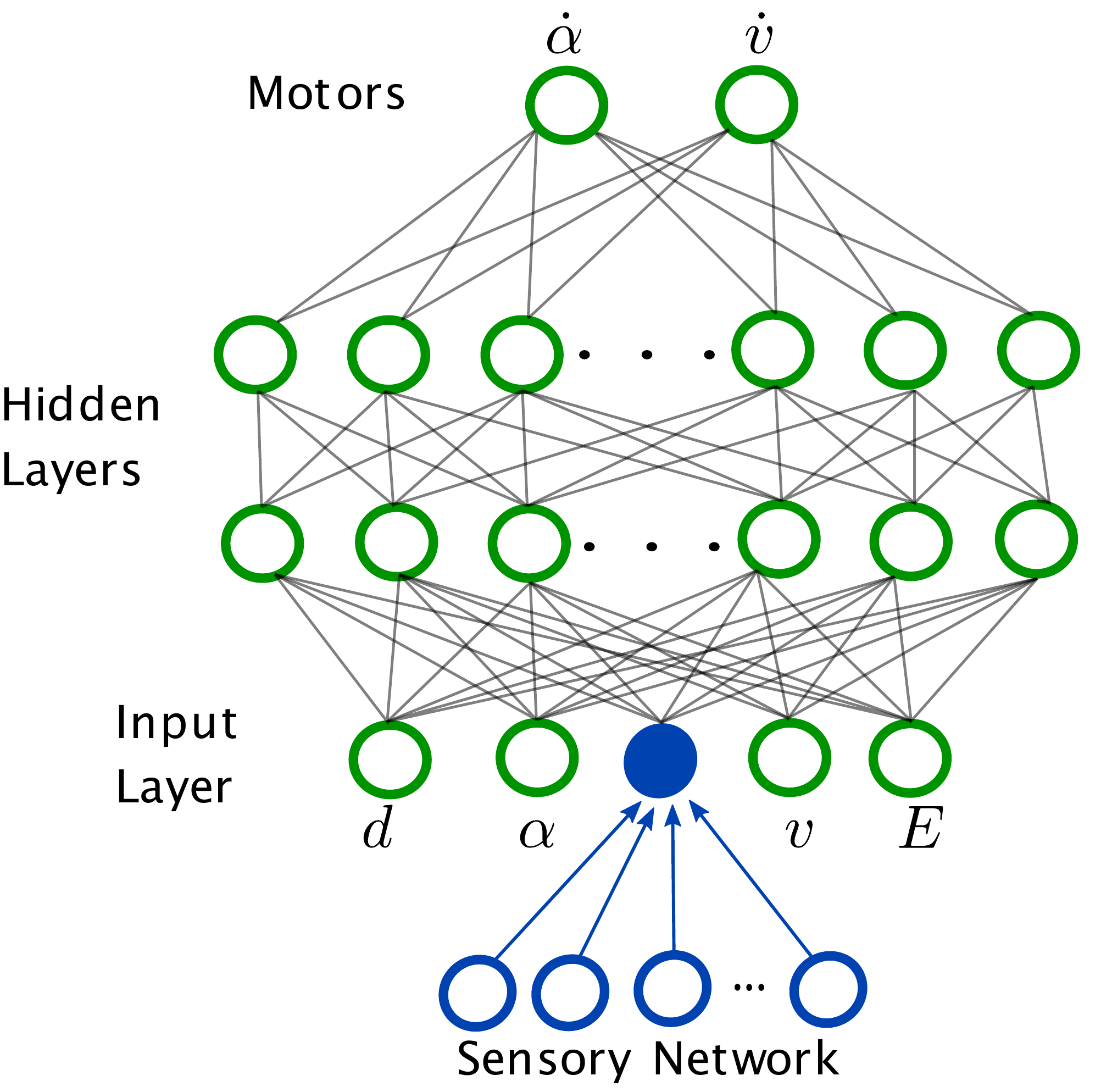}
\caption{\textit{An outline of the network controlling the foraging agent. The sensor layer receives inputs at each time step (the ingredients of the nearest food), which are processed by the plastic layer in the same way as the static sensory network, \fig\ref{fig:sensor_network}. The output of that network is given as input to the motor network, along with the distance $d$ and angle $\alpha$ to the nearest food, the current velocity  $v$, and energy $E$ of the agent. These signals are processed through two hidden layers to the final output of motor commands as the linear and angular acceleration of the agent}}
\label{fig:agent_network}
\end{center}
\end{figure}

As a next step, we incorporate the sensory network of static agents into embodied agents that can move around in an environment scattered with food.
To this end, we merge the static agent's network with a second, non-plastic motor network that is responsible for controlling the motion of the agent in the environment. Specifically, the original plastic network now provides the agent with information about the value of the nearest food. The embodied agent has additional sensors for the distance from the nearest food, the angle between the current velocity and the nearest food direction, its own velocity, and its own energy level (sum of consumed food values). These inputs are processed by two hidden layers (of 30 and 15 neurons) with $\tanh$ activation. The network's outputs are angular and linear acceleration, \fig\ref{fig:agent_network}. 

The embodied agents spawn in a 2D space with periodic boundary conditions along with a number of food particles that are selected such that the mean of the food value distribution is $\sim 0$. An agent can eat food by approaching it sufficiently closely, and each time a food particle is eaten, it is re-spawned with the same value somewhere randomly on the grid (following the setup of \cite{Khajehabdollahi2022}). After 5000 time steps, the cumulative reward of the agent (the sum of the values of all the food it consumed) is taken as its fitness. During the evolutionary optimization, the parameters for both the motor network (connections) and plastic network (learning rule parameters) are co-evolved, and so agents must simultaneously learn to move and discriminate good/bad food.

\begin{figure*}[!htb]
\centering
\includegraphics[width=\textwidth]{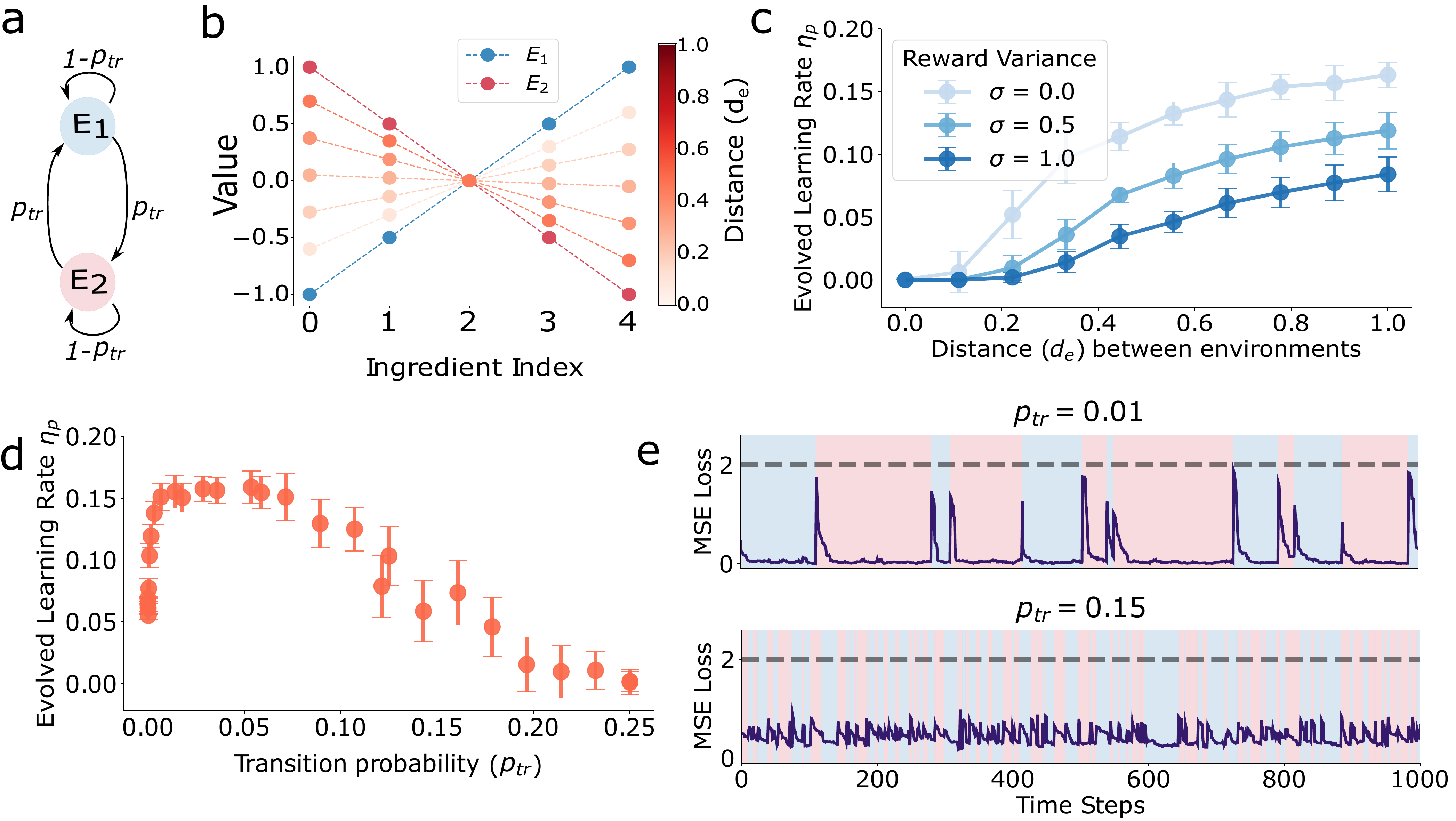}
\caption{\textit{\textbf{a.} Schematic representation of two-states Markov model with transition probability $p_{tr}$ between two environments $E_1$ and $E_2$ defined by the ingredient value distributions. \textbf{b.} We vary the $E_2$ environment by changing the ingredient values linearly $E_2 = (1- 2 d_e) E_1$, the $d_e$ is indicated by the color.
\textbf{c.} The evolved learning rate $\eta_p$ grows with the distance $d_e$ between the environments and decreases with the reward variance $\sigma$. \textbf{d.} The environment transition probability $p_{tr}$ 
(here for $d_e = 1$ and $\sigma = 0.25$) has a non-monotonous relationship with the evolved learning rate $\eta_p$. Up to a certain point, more rapid transitions lead to faster learning, but too rapid environmental transition leads to a reduction of the evolved learning rate. \textbf{e}. For slow environmental transition (top), the agent fully adapts to the environment after each transition. If the transitions happen fast (bottom), the agent maintains an intermediate position between the two environments and never fully adapts to either of them.}}
\label{fig:result_static}
\end{figure*}

\subsection{Plasticity rule parametrization} 

Reward-modulated plasticity is one of the most promising explanations for biological credit assignment \cite{Legenstein2008}. In our network, the plasticity rule that updates the weights of the linear sensor network is a reward-modulated rule which is parameterized as a linear combination of the input, the output, and the reward at each time step:
\begin{multline}\Delta W_t = \eta_p [ R_t \cdot \overbrace{ (\theta_1 X_t y_t + \theta_2 y_t  +   \theta_3 X_t + \theta_4)}^{\text{Reward Modulated}} \\ +  \underbrace{(\theta_5 X_t y_t + \theta_6 y_t +   \theta_7 X_t  + \theta_8)}_{\text{Hebbian}} ].
\end{multline}
Additionally, after each plasticity step, the weights are normalized by mean subtraction, an important step for the stabilization of Hebbian-like plasticity rules \cite{Zenke2016}.

We use a genetic algorithm to optimize the learning rate $\eta_p$ and amplitudes of different terms $ {\theta} = (\theta_1, \dots, \theta_8)$. The successful plasticity rule after many food presentations must converge to a weight vector that predicts the correct food values (or allows the agent to correctly decide whether to eat a food or avoid it). 

To have comparable results, we divide $ {\theta} = (\theta_1, \dots, \theta_8)$ by  $\theta_\mathrm{max} = \max_{k}|\theta_k|$. So that ${\theta} / \theta_\mathrm{max} = {\theta}^\mathrm{norm} \in [-1,1]^8$. We then multiply the learning rate $\eta_p$ with $\theta_\mathrm{max}$ to maintain the rule's evolved form unchanged, $\eta_p^\mathrm{norm} = \eta_p \cdot \theta_\mathrm{max}$. In the following, we always use normalized $\eta_p$ and ${\theta}$, omitting $^\mathrm{norm}$.

\subsection{Evolutionary Algorithm}
To evolve the plasticity rule and the moving agents' motor networks, we use a simple genetic algorithm with elitism \cite{Deb2011}. The agents' parameters are initialized at random (drawn from a Gaussian distribution), then the sensory network is trained by the plasticity rule and finally, the agents are evaluated. After each generation, the best-performing agents (top 10 \% of the population size) are selected and copied into the next generation. The remaining 90 \% of the generation is repopulated with mutated copies of the best-performing agents. We mutate agents by adding independent Gaussian noise ($\sigma = 0.1$) to its parameters.

\section{Results}

\subsection{Environmental and reward variability control the evolved learning rates of the static agents}
To start with, we consider a static agent whose goal is to identify the value of presented food correctly. 
The static reward-prediction network quickly evolves the parameters of the learning rule, successfully solving the prediction task. We first look at the evolved learning rate $\eta_p$, which determines how fast (if at all) the network's weight vector is updated during the lifetime of the agents. 
We identify three factors that control the learning rate parameter the EA converges to: the distance between the environments, the noisiness of the reward, and the rate of environmental transition. 

The first natural factor is the distance $d_e$ between the two environments, with a larger distance requiring a higher learning rate, \fig\ref{fig:result_static}c. 
This is an expected result since the convergence time to the  ``correct'' weights is highly dependent on the initial conditions.  
If an agent is born at a point very close to optimality, which naturally happens if the environments are similar, the distance it needs to traverse on the fitness landscape is small.  Therefore it can afford to have a small learning rate, which leads to a more stable convergence and is not affected by noise.

A second parameter that impacts the learning rate is the variance of the rewards. The reward an agent receives for the plasticity step contains a noise term $\xi$ that is drawn from a zero mean Gaussian distribution with standard deviation $\sigma$. This parameter controls the unreliability of the agent's sensory system, i.e., higher $\sigma$ means that the information the agent gets about the value of the foods it consumes cannot be fully trusted to reflect the actual value of the foods. As $\sigma$ increases, the learning rate $\eta_p$ decreases, which means that the more unreliable an environment becomes, the less an agent relies on plasticity to update its weights,  \fig\ref{fig:result_static}c. Indeed for some combinations of relatively small distance $d_e$ and high reward variance $\sigma$, the EA converges to a learning rate of $\eta_p \approx 0$. This means that the agent opts to have no adaptation during its lifetime and remain at the mean of the two environments. It is an optimal solution when the expected loss due to ignoring the environmental transitions is, on average, lower than the loss the plastic network will incur by learning via the (often misleading because of the high $\sigma$) environmental cues.

A final factor that affects the learning rate the EA will converge to is the frequency of environmental change during an agent's lifetime. Since the environmental change is modeled as a simple, two-state Markov process (\fig\ref{fig:result_static}a), the control parameter is the transition probability $p_{tr}$. 
\begin{figure*}[tb]
\centering
\includegraphics[width=0.9\textwidth]{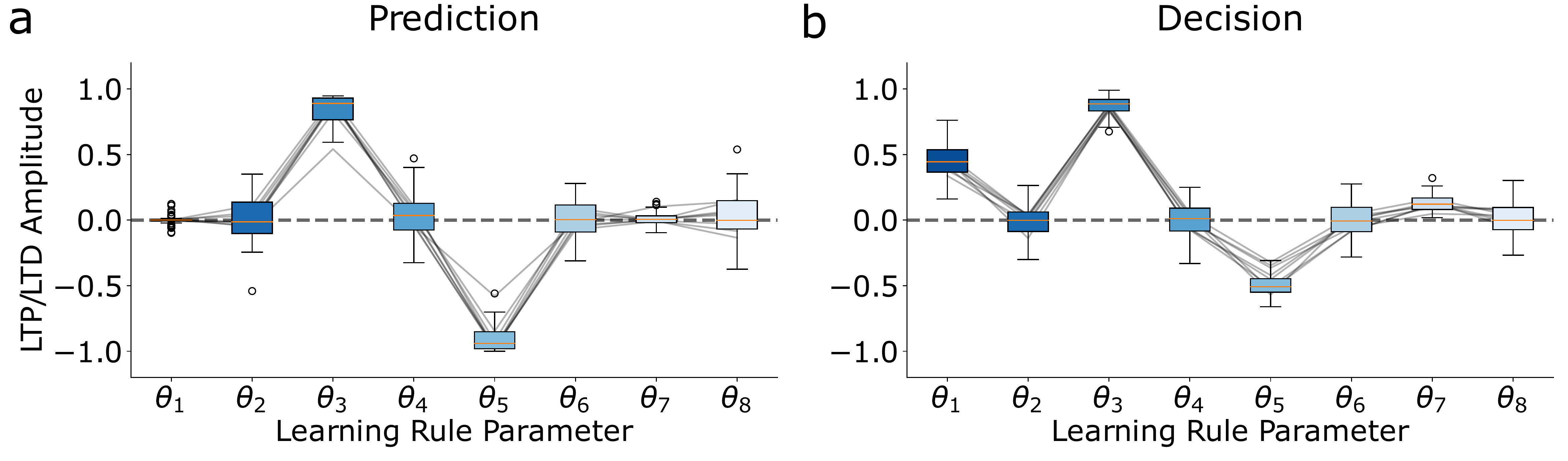}
\caption{\textit{The evolved parameters $ {\theta} = (\theta_1, \dots, \theta_8)$ of the plasticity rule for the reward prediction (\textbf{a.}) and the decision (\textbf{b.}) tasks, for a variety of parameters ($p_{tr} = 0.01$,  $d_e \in {0, 0.1, \dots, 1}$, and $\sigma \in {0, 0.1, \dots, 1}$ in all 100 combinations). Despite the relatively small difference between the tasks, the evolved learning rules differ considerably. For visual guidance, the lines connect $\theta$s from the same run.}}
\label{fig:Rules}
\end{figure*}

When keeping everything else the same, the learning rate rapidly rises as we increase the transition probability from 0, and after reaching a peak, it begins to decline slowly, eventually reaching zero (\fig\ref{fig:result_static}d). This means that when environmental transition is very rare, agents opt for a very low learning rate, allowing a slow and stable convergence to an environment-appropriate weight vector that leads to very low losses while the agent remains in that environment. As the rate of environmental transition increases, faster learning is required to speed up convergence in order to exploit the (comparatively shorter) stays in each environment. Finally, as the environmental transition becomes too fast, the agents opt for slower or even no learning, which keeps them near the middle of the two environments, ensuring that the average loss of the two environments is minimal (\fig\ref{fig:result_static}d).

\subsection{The form of the evolved learning rule depends on the task: Decision vs. Prediction}

The plasticity parameters  ${\theta} = (\theta_1, \dots, \theta_8)$ for the reward-prediction task converge on approximately the same point, regardless of the environmental parameters (\fig\ref{fig:Rules}a). In particular, $\theta_3 \to 1$, $\theta_5 \to -1$, $\theta_i \to 0$ for all other $i$, and thus the learning rule converges to:
\begin{equation}
    \Delta W_t  = \eta_p [\theta_3 X_tR_t + \theta_5 X_ty_t] \approx \eta_p X_t (R_t - y_t).
\end{equation}
Since by definition $y_t = g(W_t X_t^T) = W_t X_t^T$ ($g(x) = x$ in this experiment) and $R_t = W^c X_t^T + \xi$ we get: 
\begin{equation}
     \Delta W_t = \eta_p X_t(W^c - W_t)X_t^T + \eta_p \xi X_t^T.
\end{equation}
Thus the distribution of $\Delta W_t$ converges to a distribution with  mean 0 and variance depending on $\eta_p$ and $\sigma$ and $W$ converges to $W^c$. 
So this learning rule will match the agent's weight vector with the vector of ingredient values in the environment.

We examine the robustness of the learning rule the EA discovers by considering a slight modification of our task. Instead of predicting the expected food value, the agent now needs to decide whether to eat the presented food or not. This is done by introducing a step-function nonlinearity ($g(x) = 1$ if $x \geq 1$ and 0 otherwise). Then the output $y(t)$ is computed as:
\begin{equation}
y_t =
    \begin{cases}
        1, & \text{if  } \ W_tX_t^T \geq 0, \\
        0, & \text{if  } \ W_tX_t^T < 0. \\
    \end{cases} \label{eq:nonlineariry}
\end{equation}
Instead of the MSE loss between prediction and actual value, the fitness of the agent is now defined as the sum of the food values it chose to consume (by giving $y_t = 1$). Besides these two changes, the setup of the experiments remains exactly the same.

The qualitative relation between $\eta_p$ and parameters of environment $d_e, \sigma$ and $p_{tr}$ is preserved in the changed experiment. However, the resulting learning rule is significantly different (\fig\ref{fig:Rules}). The evolution converges to the following learning rule:
\begin{equation}
\Delta W_t =  
    \begin{cases}
        \eta_p X_t [\theta_3 R_t + \theta_7], \ y_t = 0, \\
        \eta_p X_t[(\theta_1 + \theta_3) R_t + (\theta_5 + \theta_7)], \  y_t = 1. \\
    \end{cases}
\end{equation}
In both cases, the rule has the form $ \Delta W_t =  \eta_p X_t [\alpha_y R_t + \beta_y]$. Thus, the $\Delta W_t $ is positive or negative depending on whether the reward $R_t$ is above or below a threshold ($ \gamma = -\beta_y / \alpha_y$) that depends on the output decision of the network ($y_t = 0$ or $1$). 

Both learning rules (for the reward-prediction and decision tasks) have a clear Hebbian form (coordination of pre- and post-synaptic activity) and use the incoming reward signal as a threshold. These similarities indicate some common organizing principles of reward-modulated learning rules, but their significant differences highlight the sensitivity of the optimization process to task details.

\subsection{The learning rate of embodied agents depends on environmental variability}

\begin{figure*}
\centering
\includegraphics[width=\textwidth]{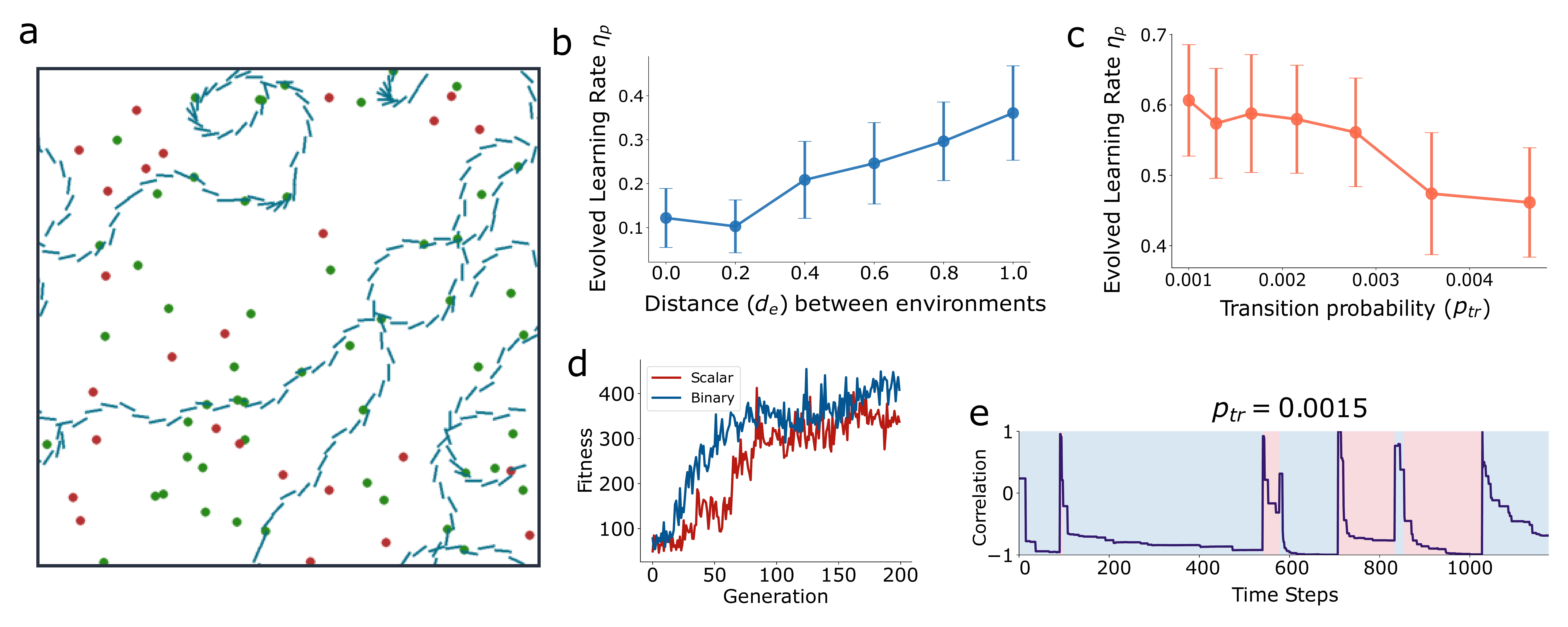}\vspace{-.8em}
\caption{\textit{\textbf{a.} The trajectory of an agent (blue line) in the 2D environment. A well-trained agent will approach and consume food with positive values (green dots) and avoid negative food (red dots). \textbf{b.} The learning rate of the plastic sensory network $eta_p$ grows with the distance between environments $d_e$ \textbf{c.} and decreases with the frequency of environmental change. \textbf{d.} The fitness of an agent (measured as the total food consumed over its lifetime) increases over generations of the EA for both the scalar and binary readouts in the sensory network. \textbf{e.} The Pearson correlation coefficient of an evolved agent's weights with the ingredient value vector of the current environment ($E_1$ - blue, $E_2$ - red). In this example, the agent's weights are anti-correlated with its environment, which is not an issue for performance since the motor network can interpret the inverted signs of food.}}
\label{fig:Foraging}
\end{figure*}

We now turn to the moving embodied agents in the 2D environment. To optimize these agents, both the motor network's connections and the sensory network's plasticity parameters evolve simultaneously. Since the motor network is initially random and the agent has to move to find food, the number of interactions an agent experiences in its lifetime can be small, slowing down the learning.
However, having the larger motor network also has benefits for evolution because it allows the output of the plastic network to be read out and transformed in different ways, resulting in a broad set of solutions.

The agents can solve the task effectively by evolving a functional motor network and a plasticity rule that converges to interpretable weights (\fig\ref{fig:Foraging}a). After $\sim$ 100 evolutionary steps (\fig\ref{fig:Foraging}d), the agents can learn the ingredient value distribution using the plastic network and reliably move towards foods with positive values while avoiding the ones with negative values.

We compare the dependence of the moving and the static agents on the parameters of the environment: $d_e$ and the state transition probability $p_{tr}$. At first, in order to simplify the experiment, we set the transition probability to $0$, but fixed the initial weights to be the average of $E_1$ and $E_2$, while the real state is $E_2$. In this experiment, the distance between states $d_e$ indicates twice the distance between the agent's initial weights and the optimal weights (the environment's ingredient values) since the agent is initialized at the mean of the two environment distributions. Same as for the static agent, the learning rate increases with the distance $d_e$  (\fig\ref{fig:Foraging}b).

Then, we examine the effect of the environmental transition probability $p_{tr}$ on the evolved learning rate $\eta_p$. In order for an agent to get sufficient exposure to each environment, we scale down the probability $p_{tr}$ from the equivalent experiment for the static agents. We find that as the probability of transition increases, the evolved learning rate $\eta_p$ decreases (\fig\ref{fig:Foraging}c). This fits with the larger trend for the static agent, although there is a clear difference when it comes to the increase for very small transition probabilities that were clearly identifiable in the static but not the moving agents. This could be due to much sparser data and possibly the insufficiently long lifetime of the moving agent (the necessity of scaling makes direct comparisons difficult). Nevertheless, overall we see that the associations observed in the static agents between environmental distance $d_e$ and transition probability $p_{tr}$ and the evolved learning rate $\eta_p$ are largely maintained in the moving agents. Still, more data would be needed to make any conclusive assertions about the exact effect of these environmental parameters on the emerging plasticity mechanisms.

\subsection{Rule redundancy in the embodied agents}
\begin{figure*}
\centering
\includegraphics[width=0.8\textwidth]{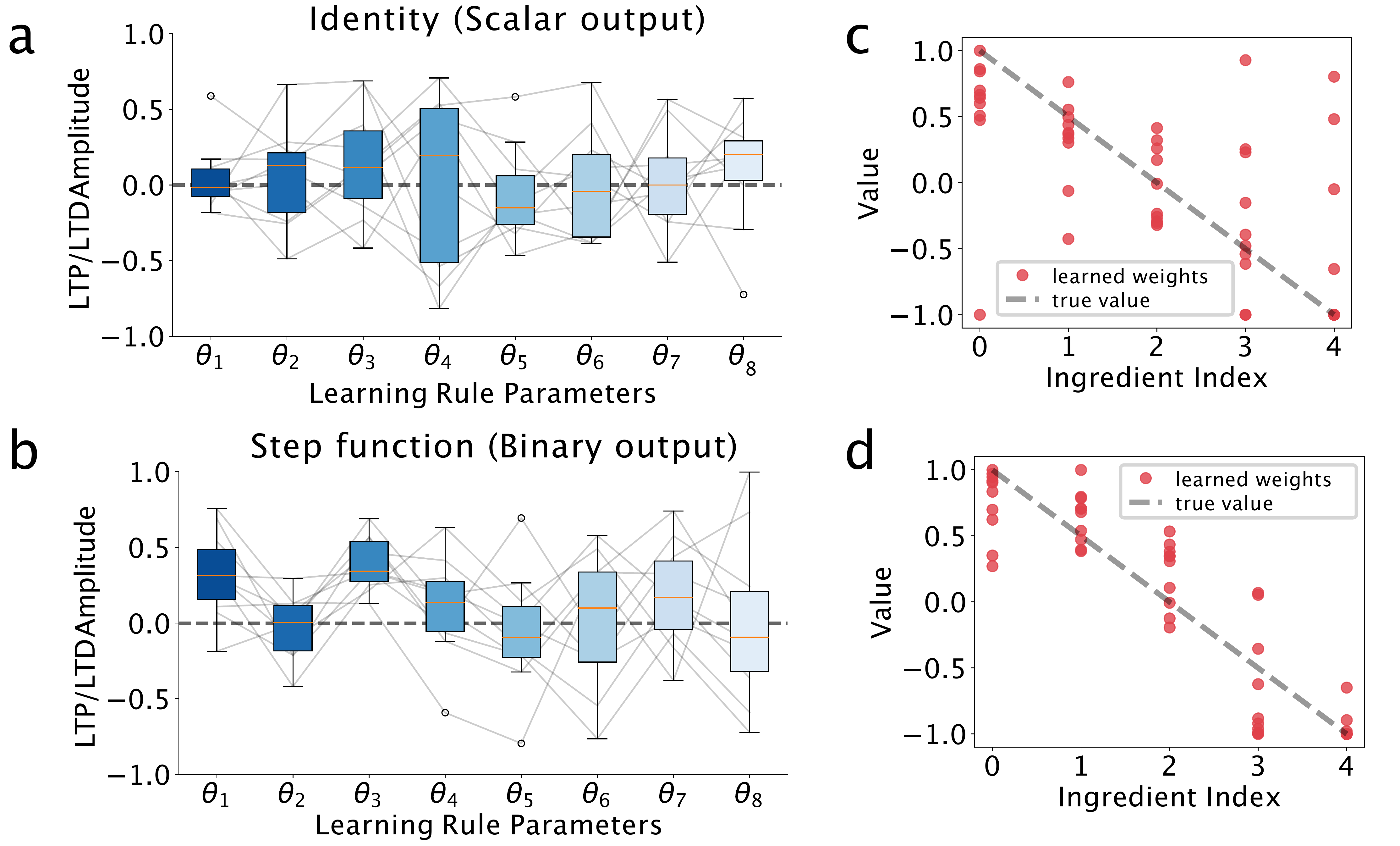}
\caption{\textit{The evolved parameters of moving agents' plasticity rule for the $g(s) = x$, identity  (\textbf{a.}) and the step function (Eq.~\ref{eq:nonlineariry}) (\textbf{b.}) sensory networks (the environmental parameters here are $d_e \in [0, 1], \ \sigma = 0$ and $p_{tr} = 0.001$). The step  function (binary output) network evolved a more structured plasticity rule (e.g., $\theta_3>0$ for all realizations) than the linear network. Moreover, the learned weights for the identity network (\textbf{c.}) have higher variance and correlate significantly less with the environment's ingredient distribution compared to the learned weights for the thresholded network (\textbf{d.})}}
\label{fig:Foraging_Rules}
\end{figure*}
A crucial difference between the static and the moving agents is the function the plasticity has to perform. While in the static agents, the plasticity has to effectively identify the exact value distribution of the environment in order to produce accurate predictions, in the embodied agents, the plasticity has to merely produce a representation of the environment that the motor network can evolve to interpret adequately enough to make decisions about which food  to consume.

To illustrate the difference, we plot the Pearson correlation coefficient between an agent's weights and the ingredient values of the environment it is moving in (\fig\ref{fig:Foraging}e). We use the correlation instead of the MSE loss (which we used for the static agents in \fig\ref{fig:result_static}e) because the amplitude of the weight vector varies a lot for different agents and meaningful conclusions cannot be drawn from the MSE loss. For many agents, the learned weights are consistently anti-correlated with the actual ingredient values (an example of such an agent is shown in \fig\ref{fig:Foraging}e). This means that the output of the sensory network will have the opposite sign from the actual food value. While in the static network, this would lead to very bad predictions and high loss, in the foraging task, these agents perform exactly as well as the ones where the weights and ingredients values are positively correlated, since the motor network can simply learn to move towards food  for which it gets a negative instead of a positive sensory input.

This additional step of the output of the plastic network going through the motor network before producing any behavior has a strong effect on the plasticity rules that the embodied agents evolve. Specifically, if we look at the emerging rules the top performing agents have evolved (\fig\ref{fig:Foraging_Rules}a), it becomes clear that, unlike the very well-structured rules of the static agents (\fig\ref{fig:Rules}a), there is now virtually no discernible pattern or structure. The difference becomes even clearer if we look at the learned weights (at the end of a simulation) of the best-performing agents (\fig\ref{fig:Foraging_Rules}c). While there is some correlation with the environment's ingredient value distribution, the variance is very large, and they do not seem to converge on the  ``correct'' values in any way. This is to some extent expected since, unlike the static agents where the network's output has to be exactly correct, driving the evolution of rules that converge to the precise environmental distribution, in the embodied networks, the bulk of the processing is done by the motor network which can evolve to interpret the scalar value of the sensory network's output in a variety of ways. Thus, as long as the sensory network's plasticity rule co-evolves with the motor network, any plasticity rule that learns to produce consistent information about the value of encountered food  can potentially be selected.

To further test this assumption, we introduce a bottleneck of information propagation between the sensory and motor networks by using a step-function nonlinearity on the output of the sensory network (Eq.~\ref{eq:nonlineariry}). Similarly to the decision task of the static network, the output of the sensory network now becomes binary. This effectively reduces the flow of information from the sensory to the motor network, forcing the sensory network to consistently decide whether food should be consumed (with the caveat that the motor network can still interpret the binary sign in either of two ways, either consuming food  marked with $1$ or the ones marked with $0$ by the sensory network). The agents perform equally well in this variation of the task as before  (\fig\ref{fig:Foraging}d), but now, the evolved plasticity rules seem to be more structured (\fig\ref{fig:Foraging_Rules}b). Moreover, the variance of the learned weights in the best-performing agents is significantly reduced (\fig\ref{fig:Foraging_Rules}d), which indicates that the bottleneck in 
 the sensory network is increasing selection pressure for rules that learn the environment's food distribution  accurately.
 
\section{Discussion}

We find that different sources of variability have a strong impact on the extent to which evolving agents will develop neuronal plasticity mechanisms for adapting to their environment. A diverse environment, a reliable sensory system, and a rate of environmental change that is neither too large nor too small are necessary conditions for an agent to be able to effectively adapt via synaptic plasticity. Additionally, we find that minor variations of the task an agent has to solve or the parametrization of the network can give rise to significantly different plasticity rules. 

Our results partially extend to embodied artificial agents performing a foraging task. We show that environmental variability also pushes the development of plasticity in such agents. Still, in contrast to the static agents, we find that the interaction of a static motor network with a plastic sensory network gives rise to a much greater variety of well-functioning learning rules. We propose a potential cause of this degeneracy; as the relatively complex motor network is allowed to read out and process the outputs from the plastic network, any consistent information coming out of these outputs can be potentially interpreted in a behaviorally useful way. Reducing the information the motor network can extract from the sensory system significantly limits learning rule variability.

Our findings on the effect of environmental variability concur with the findings of previous studies \cite{Tjarko2020} that have identified  the constraints that environmental variability places on the evolutionary viability of learning behaviors. We extend these findings in a mechanistic model which uses a biologically plausible learning mechanism (synaptic plasticity). We show how a simple evolutionary algorithm can optimize the different parameters of a simple reward-modulated plasticity rule for solving simple prediction and decision tasks. Reward-modulated plasticity has been extensively studied as a plausible mechanism for credit assignment in the brain \cite{Razvan2007, Baras2007, Legenstein2008}  and has found several applications in artificial intelligence and robotics tasks \cite{Burms2015, Bing2019}. Here, we demonstrate how such rules can be very well-tuned to take into account different environmental parameters and produce optimal behavior in simple systems.

Additionally, we demonstrate how the co-evolution of plasticity and static functional connectivity in different sub-networks fundamentally changes the evolutionary pressures on the resulting plasticity rules, allowing for greater diversity in the form of the learning rule and the resulting learned connectivity. Several studies have demonstrated how, in biological networks, synaptic plasticity heavily interacts with \cite{Butz2014, Bassi2019, Bernaez2022} and is driven by network topology \cite{Giannakakis2023}. Moreover, it has been recently demonstrated that biological plasticity mechanisms are highly redundant in the sense that any observed neural connectivity or recorded activity can be achieved with a variety of distinct, unrelated learning rules
\cite{Ramesh2023}. This observed redundancy of learning rules in biological settings complements our results and suggests that the function of plasticity rules cannot be studied independently of the connectivity and topology of the networks they are acting on. 

The optimization of functional plasticity in neural networks is a promising research direction both as a means to understand biological learning processes and as a tool for building more autonomous artificial systems. Our results suggest that reward-modulated plasticity is highly adaptable to different environments and can be incorporated into larger systems that solve complex tasks.

\section{Future work}

This work studies a simplified toy model of neural network learning in stochastic environments. Future work could be built on this basic framework to examine more complex reward distributions and sources of environmental variability. Moreover, a greater degree of biological realism could be added by studying more plausible network architectures (multiple plastic layers, recurrent and feedback connections) and more sophisticated plasticity rule parametrizations. 

Additionally, our foraging simulations were constrained by limited computational resources and were far from exhaustive. Further experiments can investigate environments with different constraints, food distributions, multiple seasons, more complex motor control systems and interactions of those systems with different sensory networks as well as the inclusion of plasticity on the motor parts of the artificial organisms. 

 \section{Acknowledgements}
This work was supported by a Sofja Kovalevskaja Award from the Alexander von Humboldt Foundation. EG and SK thank the International Max Planck Research School for Intelligent Systems (IMPRS-IS) for their support. We acknowledge the support from the BMBF through the T\"ubingen AI Center (FKZ: 01IS18039A). AL is a member of the Machine Learning Cluster of Excellence, EXC number 2064/1 – Project number 39072764.

\footnotesize
\bibliographystyle{apalike}
\bibliography{main}

\end{document}